\documentstyle [12pt]{article}

\begin{document}
\input psfig
\title{
\begin{flushright}
CLNS 95/1363 \\
CLEO 95-18 \\
\end{flushright}
\vskip 1.4in
\begin{center}
{\bf Measurement of the Inclusive Semi-electronic $D^0$ Branching
Fraction}
\end{center}
}
\author{}
\maketitle

\begin{center}
Y.~Kubota, M.~Lattery, M.~Momayezi, J.K.~Nelson, S.~Patton,
R.~Poling, T.~Riehle, V.~Savinov, and R.~Wang \\

{\it University of Minnesota, Minneapolis, Minnesota 55455} \\
\vskip 0.25in

M.S.~Alam, I.J.~Kim, Z.~Ling, A.H.~Mahmood, J.J.~O'Neill,
H.~Severini, C.R.~Sun, S.~Timm, and F.~Wappler \\

{\it State University of New York at Albany, Albany, New York 12222} \\
 \vskip 0.25in

G.~Crawford, J.E.~Duboscq, R.~Fulton, D.~Fujino, K.K.~Gan,
K.~Honscheid, H.~Kagan, R.~Kass, J.~Lee, M.~Sung, C.~White,
R.~Wanke, A.~Wolf, and M.M.~Zoeller  \\

{\it Ohio State University, Columbus, Ohio, 43210 }\\
\vskip 0.25in

X.~Fu, B.~Nemati, W.R.~Ross, P.~Skubic, and M.~Wood \\

{\it University of Oklahoma, Norman, Oklahoma 73019 }\\
\vskip 0.25in

M.~Bishai, J.~Fast, E.~Gerndt, J.W.~Hinson, T.~Miao, D.H.~Miller,
M.~Modesitt, E.I.~Shibata, I.P.J.~Shipsey, and P.N.~Wang \\

{\it Purdue University, West Lafayette, Indiana 47907 }\\
 \vskip 0.25in

L.~Gibbons, S.D.~Johnson, Y.~Kwon, S.~Roberts, and E.H.~Thorndike \\

{\it University of Rochester, Rochester, New York 14627  } \\
\vskip 0.25in

T.E.~Coan, J.~Dominick, V.~Fadeyev, I.~Korolkov, M.~Lambrecht,
S.~Sanghera, V.~Shelkov, R.~Stroynowski,
I.~Volobouev, and G.~Wei  \\

{\it Southern Methodist University, Dallas, Texas 75275 }\\
\vskip 0.25in

M.~Artuso, M.~Gao, M.~Goldberg, D.~He, N.~Horwitz, S.~Kopp,
G.C.~Moneti, R.~Mountain, F.~Muheim, Y.~Mukhin, S.~Playfer,
S.~Stone, and X.~Xing \\

{\it Syracuse University, Syracuse, New York 13244} \\
 \vskip 0.25in

J.~Bartelt, S.E.~Csorna, V.~Jain, and S.~Marka \\

{\it Vanderbilt University, Nashville, Tennessee 37235} \\
 \vskip 0.25in

D.~Gibaut, K.~Kinoshita, P.~Pomianowski, and S.~Schrenk \\

{\it Virginia Polytechnic Institute and State University,
Blacksburg, Virginia, 24061 }\\
 \vskip 0.25in

B.~Barish, M.~Chadha, S.~Chan, D.F.~Cowen, G.~Eigen, J.S.~Miller,
C.~O'Grady, J.~Urheim, A.J.~Weinstein, and F.~W\"urthwein \\

{\it California Institute of Technology, Pasadena, California 91125 }\\
 \vskip 0.25in

D.M.~Asner, M.~Athanas, D.W.~Bliss, W.S.~Brower, G.~Masek,
and H.P.~Paar \\

{\it University of California, San Diego, La Jolla, California 92093 }\\
 \vskip 0.25in

J.~Gronberg, C.M.~Korte, R.~Kutschke, S.~Menary, R.J.~Morrison,
S.~Nakanishi, H.N.~Nelson, T.K.~Nelson, C.~Qiao, J.D.~Richman,
D.~Roberts, A.~Ryd, H.~Tajima, and M.S.~Witherell \\

{\it University of California, Santa Barbara, California 93106 }\\
 \vskip 0.25in

R.~Balest, K.~Cho, W.T.~Ford, M.~Lohner, H.~Park, P.~Rankin,
J.~Roy, and J.G.~Smith               \\

{\it University of Colorado, Boulder, Colorado 80309-0390 }\\
 \vskip 0.25in

J.P.~Alexander, C.~Bebek, B.E.~Berger, K.~Berkelman, K.~Bloom,
T.E.~Browder,%
\thanks{Permanent author: University of Hawaii at Manoa}
D.G.~Cassel, H.A.~Cho, D.M.~Coffman, D.S.~Crowcroft, M.~Dickson,
P.S.~Drell, D.J.~Dumas, R.~Ehrlich, R.~Elia, P.~Gaidarev,
B.~Gittelman, S.W.~Gray, D.L.~Hartill,
B.K.~Heltsley, S.~Henderson, C.D.~Jones, S.L.~Jones,
J.~Kandaswamy, N.~Katayama, P.C.~Kim, D.L.~Kreinick, T.~Lee,
Y.~Liu, G.S.~Ludwig, J.~Masui, J.~Mevissen, N.B.~Mistry, C.R.~Ng,
E.~Nordberg, J.R.~Patterson, D.~Peterson, D.~Riley, A.~Soffer, and C.~Ward  \\

{\it  Cornell University, Ithaca, New York 14853 }   \\
\vskip 0.25in

P.~Avery, A.~Freyberger, K.~Lingel, C.~Prescott, J.~Rodriguez,
S.~Yang, and J.~Yelton               \\

{\it University of Florida, Gainesville, Florida 32611 }  \\
\vskip 0.25in

G.~Brandenburg, D.~Cinabro, T.~Liu, M.~Saulnier, R.~Wilson,
and H.~Yamamoto                 \\

{\it Harvard University, Cambridge, Massachusetts 02138 }    \\
\vskip 0.25in

T.~Bergfeld, B.I.~Eisenstein, J.~Ernst, G.E.~Gladding,
G.D.~Gollin, M.~Palmer, M.~Selen, and J.J.~Thaler   \\

{\it University of Illinois, Champaign-Urbana, Illinois, 61801  }  \\
\vskip 0.25in

K.W.~Edwards, K.W.~McLean, and M.~Ogg             \\

{\it Carleton University, Ottawa, Ontario K1S 5B6
and the Institute of Particle Physics, Canada }    \\
\vskip 0.25in

A.~Bellerive, D.I.~Britton, E.R.F.~Hyatt, R.~Janicek,
D.B.~MacFarlane, P.M.~Patel, and B.~Spaan        \\

{\it McGill University, Montr\'eal, Qu\'ebec H3A 2T8
and the Institute of Particle Physics, Canada}    \\
 \vskip 0.25in

A.J.~Sadoff                     \\

{\it Ithaca College, Ithaca, New York 14850 }     \\
 \vskip 0.25in

R.~Ammar, P.~Baringer, A.~Bean, D.~Besson, D.~Coppage, N.~Copty,
R.~Davis, N.~Hancock, S.~Kotov, I.~Kravchenko, and N.~Kwak  \\

{\it University of Kansas, Lawrence, Kansas 66045}   \\
\vskip 0.25in
(CLEO Collaboration) \\
  \vskip 0.25in
\end{center}

%
\newpage
\begin{abstract}
Using the angular correlation between the $\pi^+$ emitted in
a $D^{*+} \rightarrow D^0 \pi^+$ decay and the
$e^+$ emitted in the subsequent $D^0 \rightarrow Xe^+\nu$
decay, we have measured the branching fraction for the inclusive
semi-electronic decay of the $D^0$ meson to be:
\begin{eqnarray}
{\cal B}(D^0 \rightarrow X e^+ \nu) =
[6.64 \pm 0.18 (stat.) \pm 0.29 (syst.)] \%. \nonumber
\end{eqnarray}  The result is based on
1.7 fb$^{-1}$ of $e^+e^-$ collisions recorded by the CLEO II detector
located at the Cornell Electron Storage Ring (CESR).
Combining the analysis presented in this paper with previous CLEO results
we find,
\begin{eqnarray}
\frac{{\cal B} (D^0 \rightarrow X e^+ \nu)}
     {{\cal B} (D^0 \rightarrow K^- \pi^+)} =
  1.684 \pm 0.056 (stat.) \pm 0.093(syst.) \nonumber
\end{eqnarray}
and
\begin{eqnarray}
\frac{{\cal B}(D\rightarrow K^-e^+\nu)}
     {{\cal B}(D\rightarrow Xe^+\nu)} =
0.581 \pm 0.023 (stat.) \pm 0.028(syst.). \nonumber
\end{eqnarray}
The
difference between the inclusive rate and the sum of the
measured exclusive branching fractions (measured at CLEO and other experiments)
is
$(3.3 \pm 7.2) \%$ of the inclusive rate.
\end{abstract}
\bigskip
\section{Introduction}
In this paper,  we present a new  measurement  of the inclusive
semi-electronic branching fraction
of the $D^0$ meson.
The comparison of the measured inclusive semi-leptonic branching fraction
with the sum of the observed exclusive semi-leptonic branching
fraction provides a measure of {\it missing} or unobserved modes.
Recent experimental progress on exclusive measurements has yielded precise
measurements of the dominant Cabibbo favored modes,
observation and measurement of the Cabibbo suppressed branching fractions
and stringent
upper limits on suppressed Cabibbo favored branching fractions, but has
not yielded an improvement in the measured inclusive semi-leptonic branching
fraction.
The inclusive branching fraction
measurement presented here and previous measurements of
exclusive branching fractions allows for more accurate comparison
than previously performed.
For a complete review of experimental
and theoretical developments we refer the reader to  recent reviews
{}~\cite{jeffandpat,pdg}.

In addition, we combine the inclusive result
presented here with previous CLEO results on ${\cal B}(D^0 \rightarrow K^-\pi)$
and ${\cal B}(D^0 \rightarrow K^-e^+\nu)/{\cal B}(D^0 \rightarrow K^-\pi^+)$
{}~\cite{cleo_kpi_93,cleo_93}
to obtain the ratio,
${\cal B}(D^0 \rightarrow K^-e^+\nu)/{\cal B}(D^0 \rightarrow X e^+\nu)$.
As a check of the method the observed inclusive
electron momentum spectrum is
also extracted from the data and
compared with a Monte Carlo simulation.

\section{Analysis Technique and Event Selection}
The technique to measure the absolute
inclusive semi-electronic branching fraction
of $D^0$ mesons is similar to the previous CLEO absolute
branching fraction measurement of  $D^0 \rightarrow K^-\pi^+$
\cite{cleo_kpi_93}.  Common to both analyses is the method
used to determine the number of
$D^{*+} \rightarrow D^0 \pi^+$ decays in the data with minimal
systematic bias.  It is based on the
unique two body kinematics of the $D^{*+} \rightarrow D^0 \pi^+$
decay and the topology
of $e^+ e^- \rightarrow c \bar{c}$ reactions at a center of mass energy
of $10.5$ GeV.
Briefly, the
idea is that the thrust axis
(defined to be that axis along which the projected momentum is a maximum)
for the event approximates
the $D^{*+}$ direction in the lab.
The limited amount of available phase space in
the $D^{*+} \rightarrow D^0\pi^+$
decay, results in a small  angle
between the thrust axis and the charged pion.  We denote this
angle between the thrust axis and the charged pion as $\alpha$.
Also, the magnitude of the pion momentum is correlated to the parent $D^{*+}$
momentum.  Pions with momentum greater than
225 MeV/c are kinematically forbidden
to come from the $\Upsilon(4S) \rightarrow B {\bar B}$,
${\bar B} \rightarrow D^{*+} X$,
$D^{*+} \rightarrow D^0 \pi^+$ decay chain.
This selection assures that the
$D^{*+}$ is from $e^+ e^- \rightarrow c {\bar c}$ production and the
event has a well defined thrust axis.  The top
plot in Figure~\ref{figure:sinalpha_summed}
shows the $\sin^2\alpha$ distribution for all pions
with momentum between 225 and 425 MeV/c in the data.
The peaking at low $\sin^2\alpha$ is evidence for
$D^{*+} \rightarrow D^0 \pi^+$ decays.
The total number,
$N(D^{*+} \rightarrow D^0 \pi^+)$, of decays in the sample is
$165658 \pm 1149(stat.) \pm 2485 (syst.)$. This total is identical
to that presented in Ref.~\cite{cleo_kpi_93}, as the same data and
selection criteria are used in both analyses.

The total number  of semi-electronic decays,
$N(D^{*+} \rightarrow D^0 \pi^+$, $D^0 \rightarrow Xe^+\nu)$,
is determined
by identifying an $e^+$ within a cone around the $\pi^+$ direction and
plotting the $\sin^2\alpha$ distribution for those $\pi^+$ with an
associated $e^+$.
This is achieved
by  studying  the sign correlated $\pi e$ combinations
in the data.  ``Right sign'' combinations, $\pi^+ e^+$, provide the
signal distribution and ``wrong sign'' combinations, $\pi^+ e^-$, are
studied to aid background determinations.  Once the number of
$D^0 \rightarrow X e^+ \nu$ decays has been determined, the branching
fraction is then,
\begin{eqnarray}
B(D^0\rightarrow Xe^+\nu) & = &
\frac{N(D^{*+}\rightarrow D^0\pi^+, D^0 \rightarrow Xe^+\nu)}
     {N(D^{*+}\rightarrow D^0\pi^+) \times \epsilon(D^0 \rightarrow Xe^+\nu)}
\end{eqnarray}
where $\epsilon({D^0 \rightarrow Xe^+\nu})$ is the efficiency for detecting the
electron.

A detailed description of the CLEO II detector
can be found in Ref.~\cite{cleo_nim}.
Electrons and positrons~\cite{conjugate}
are identified principally from the ratio of the
energy measured by the CsI calorimeter and the momentum measured by the
drift chamber (E/p).
Additional information on energy loss in the drift chamber
and shower shape in the calorimeter is also used to maximize the identification
efficiency and minimize the mis-identification of hadronic tracks.  The
electrons are required to have momentum greater than 0.7 GeV/c and
a polar angle with respect to the beam axis ($\theta$)
between 45$^o$ and 135$^o$, to insure
a well determined efficiency and minimal uncertainty due to mis-identified
hadronic tracks.
Furthermore it is important to
reduce the number of
electrons from $D^0 \rightarrow X \pi^0$; $\pi^0 \rightarrow
e^+e^-\gamma$, where the $e^+e^- \gamma$ final state is due to either a
Dalitz decay of the $\pi^0$ or a $\gamma$ conversion in the detector material.
This is accomplished by requiring that the identified electron, when combined
with each opposite sign track in the event,
does not yield an electron-positron
mass below 0.050 GeV/c$^2$.
Every opposite sign track is used to form these pairs, whether or not
it is identified as an electron.

In order to correlate the $\pi^+$ with
an $e^+$ a fiducial angle cut is applied in the lab frame.
We require that
$\cos(\Theta_{e - \pi}) > 0.8$,
where $\Theta_{e-\pi}$ is the angle between the $\pi$ emitted
in the initial $D^{*+} \rightarrow D^0 \pi^+$ decay and the
electron from the subsequent $D^0 \rightarrow Xe^+\nu$ decay.
The bottom histogram in Figure~\ref{figure:sinalpha_summed}
shows the $\sin^2\alpha$ distributions for $\pi$'s
after requiring an electron within this angular region;
the
solid squares are for $\pi^+e^+$ combinations (right sign)
 and the open squares are
for $\pi^+ e^-$ combinations(wrong sign).

\begin{figure}
\vspace{6.0in}
 \caption{The inclusive $\sin^2\alpha$ distribution for
candidate pions (open circles) and the derived non-$D^{*+}$
background (solid line) in the top plot.
Requiring an electron near the pion
with the same (opposite) sign
results in the solid (open) squares in the bottom plot.
  \label{figure:sinalpha_summed}}
\end{figure}

\section{Extraction of yields}
As previously stated, the yield of $D^{*+} \rightarrow D^0 \pi^+$ decays
is identical to that presented in Ref.~\cite{cleo_kpi_93}.   In this section
we detail the determination of the number of
$D^0 \rightarrow Xe^+\nu$ decays associated
with the initial $D^{*+} \rightarrow D^0 \pi^+$ decay.

The $\sin^2\alpha$ distribution for $\pi^+e^+$ (right sign)
combinations contains three distinct components: signal
and two types of background.
One background has a $\sin^2\alpha$
distribution that is identical to the signal
as it originates from the decay
$D^{*+} \rightarrow D^0 \pi^+$, $D^0 \rightarrow X f_{\pi^+e^+}$, where
$f_{\pi^+e^+}$ denotes a $e^+$ from either a hadronic track
mis-identified as a electron or an electron from a $\pi^0 \rightarrow
e^+e^-\gamma$ final state.  The other background is due to
random soft pions (225 to 425 MeV/c in momentum), in coincidence
with
an electron, and is not as
sharply peaked near $\sin^2\alpha = 0$ as the signal distribution.

The $\sin^2\alpha$ distribution for $\pi^+ e^-$ (wrong sign)
combinations is devoid of signal but contains the same two sources
of background as the right sign distribution~\cite{rareD}.
The shapes for
these backgrounds in the right sign and wrong sign
distributions are identical, although the normalizations
differ.
This difference in normalization is
the result of the hadronic track mixture
($\pi/K$ ratio)  combined with
the hadronic track mis-identification
rates.
For the non-$D^{*+}$
pion  background, the normalization is
different due to charge conservation in the event.

To use as much information as possible, the right sign and
wrong sign distributions are fit simultaneously to the
following functional forms:
\begin{eqnarray}
{\cal G}_{rs}(p_\pi,\sin^2\alpha)
                    & = & N_{D^0 \rightarrow Xe^+\nu}(p_\pi) \nonumber
                         g_{D^0 \rightarrow Xe^+\nu}(\sin^2\alpha,p_\pi) + \\
                     &&  B_{rs}(p_\pi)P_2(\sin^2\alpha)
\end{eqnarray}
\begin{eqnarray}
{\cal G}_{ws}(p_\pi,\sin^2\alpha)
                     & = & N_{D^0 \rightarrow X f_e}(p_\pi) \nonumber
                         g_{D^0 \rightarrow X f_e}(\sin^2\alpha,p_\pi) + \\
                     &&  B_{ws}(p_\pi)P_2(\sin^2\alpha).
\end{eqnarray}

The  expected $\sin^2\alpha$ distributions,
$g_{D^0 \rightarrow Xe^+\nu}(\sin^2\alpha,p_\pi)$
and $g_{D^0 \rightarrow X f_e}(\sin^2\alpha,p_\pi)$, for the right sign
and wrong sign distributions are obtained from a
Monte Carlo simulation.
The Monte Carlo simulation correctly reproduces the measured $D^{*+}$
production momentum distribution, and simulates $D^0 \rightarrow Xe^+\nu$
decays via the ``cocktail'' of exclusive modes presented in
Appendix A.
The second order polynomial, $P_2$,
is constrained to
have the same shape for both the wrong and right sign distributions.
The yield of $D^0 \rightarrow Xe^+\nu$ decays,
$N_{D^0 \rightarrow Xe^+\nu}(p_\pi)$,
the yield of mis-identified hadrons and electrons
from $\pi^0 \rightarrow e^+e^-\gamma$ in the wrong sign distribution,
$N_{D^0 \rightarrow X f_e}(p_\pi)$, the normalizations and shape,
$B_{rs}(p_\pi)$,  $B_{ws}(p_\pi)$ and $P_2$, of the background polynomial  are
determined from the fits to the $\sin^2\alpha$ distribution
of the right sign and wrong sign samples in bins of pion momentum, $p_\pi$.

The $\sin^2\alpha$ distributions for the data, with the resulting
fits
overlaid, are shown in Figures~\ref{figure:data1} and~\ref{figure:data2}
and the right sign and wrong sign yields are presented
in Table~\ref{table:yields}.  The right sign
yields reported in this table have
a contribution due to $D^0 \rightarrow Xf_{\pi^+e^+}$
backgrounds.

\begin{table}
\begin{center}
\begin{tabular}{lcc}
$p(\pi)$    & \multicolumn{2}{c}{Yields} \\
(MeV/c)     & Right Sign & Wrong Sign \\ \hline
225 - 250 &$1232\pm 53$& $ 32\pm 31$\\
250 - 275 &$1071\pm 49$& $ 74\pm 29$\\
275 - 300 &$ 935\pm 44$& $ 45\pm 25$\\
300 - 325 &$ 689\pm 38$& $ 39\pm 22$\\
325 - 350 &$ 414\pm 32$& $-29\pm 18$\\
350 - 375 &$ 259\pm 25$& $ 36\pm 17$\\
375 - 400 &$ 166\pm 20$& $ -4\pm 12$\\
400 - 425 &$  79\pm 15$& $  0\pm 11$\\ \hline
Total         &$4845\pm 104$&$193\pm 62$ \\ \hline
\end{tabular}
\caption{The total yield of right sign and wrong sign events as a function
of pion momentum.  Backgrounds have not yet been
subtracted.\label{table:yields}
}
\end{center}
\end{table}
\begin{figure}
\vspace{6.0in}
 \caption{The $\sin^2\alpha$ distribution for pions with momentum
between 225 and 325 MeV/c  with an identified electron
within $\cos\Theta_{\pi-e} > 0.8$.
Events
with the electron and pion having the same sign (right sign) are
plotted on the left side, the opposite sign events (wrong sign)
are plotted on the right side.  The points represent the data and
the histogram is the result of the fit.  The dashed line represents
the random pion-electron background and is modeled by a second
order polynomial.
  \label{figure:data1}}
\end{figure}
\begin{figure}[p]
\vspace{6.0in}
 \caption{The $\sin^2\alpha$ distribution for pions with momentum
between 325 and 425 MeV/c  with an identified electron
within $\cos\Theta_{\pi-e} > 0.8$.
Events
with the electron and pion having the same sign (right sign) are
plotted between on the left side, the opposite sign events (wrong sign)
are plotted between on the right side.  The points represent the data and
the histogram is the result of the fit.  The dashed line represents
the random pion-electron background and is modeled by a second
order polynomial.
  \label{figure:data2}}
\end{figure}

\section{Determination of the background contribution to the signal}

In this section the magnitude of the $f_{\pi^+e^+}$ background to
the right sign signal yield is determined.  The two contributions
to this background are the following decay chains:
$D^{*+}\rightarrow D^0 \pi^+$, $D^0 \rightarrow X h^+$,  where
the $h^+$ is a hadronic track mis-identified
as an electron and
$D^{*+}\rightarrow D^0 \pi^+$, $D^0 \rightarrow X \pi^0$,
$\pi^0\rightarrow e^+e^- \gamma$.  We denote this sum for the
right sign background as
\begin{eqnarray}
N_{rs} & = & N(X \pi^0)
                   f_{\pi^+e^+}(X \pi^0) +
                   N(X h^+)
                   f_{\pi^+e^+}(X h^+),
                \label{eq:rs_sum}
\end{eqnarray}
where $N(X \pi^0)$ $[N(X h^+)]$
is the number of inclusive $D^0 \rightarrow X \pi^0$
[$D^0 \rightarrow X h^+$] decays in the data and
$f_{\pi^+e^+}(X \pi^0)$ [$f_{\pi^+e^+}(X h^+)$]
is the efficiency for detecting this background as signal.
We can define the same sum for the wrong sign yield as
\begin{eqnarray}
N_{ws} & = & N(X \pi^0)
                   f_{\pi^+e^-}(X \pi^0) +  N(X h^-)
                   f_{\pi^+e^-}(X h^-).
\label{eq:ws_sum}
\end{eqnarray}
The only difference between the wrong sign yield and the right sign
background is due to the fact that the positive tracks from $D^0$ decays
are much less likely to be kaons than negative tracks from $D^0$ decays.
Using
$   f_{\pi^+e^-}(X \pi^0)  =
    f_{\pi^+e^+}(X \pi^0)$, we find
\begin{eqnarray}
N_{rs} & = & N_{ws} -
                N(X h^-)
                f_{\pi^+e^-}(X h^-) +
                N(X h^+)
                f_{\pi^+e^+}(X h^+).
\end{eqnarray}
If $  N(X h^+)
      f_{\pi^+e^+}(X h^+) =
      N(X h^-)
      f_{\pi^+e^-}(X h^-)$,
then the wrong sign yield would be equal to the background contribution
to the right sign yield.  However, the $\pi^+:K^+$ ratio of
$h^+$ tracks originating from $D^0$ mesons is quite different from the
$\pi^-:K^-$ ratio.
Using world averages ~\cite{pdg} of the measured
$D^0$ branching fractions, the $\pi^+$:$K^+$ ratio is 96:4
while the $\pi^-$:$K^-$ ratio is 42:58, for pions and kaons
from $D^0$ mesons which pass the same geometry and momentum  criteria
as for electrons.
This difference coupled
with different mis-identification rates for $\pi$'s and $K$'s leads
to a small correction to the wrong sign yield.

The probability for a $\pi^+$ track to be mis-identified
as a $e^+$ is determined by studying a large sample
$K^0_s \rightarrow \pi^+ \pi^-$ decays in the data.  This sample
is large enough to determine the mis-identification probability for
charged pions as a function of their momentum.  This
probability is measured to be $(0.056 \pm 0.015) \%$ for pions
with momentum between 0.7 and 0.9 GeV/c.  It
rises as a function of pion momentum, such that for
pions with momentum between 1.9 and 2.5 GeV/c, it
is measured to be $(0.250 \pm 0.059) \%$. Convoluting the
momentum dependent mis-identification probability with a Monte
Carlo simulation of the $\pi^+$ momentum distribution from
$D^0$ and $\bar{D^0}$ decays, we find the mis-identification
probability integrated over pion momentum to be $(0.102 \pm 0.016) \%$
for the right sign pions and $(0.093 \pm 0.011) \%$ for the
wrong sign pions.  These numbers differ due to the different
momentum spectrum for right sign and wrong sign pions.
The error is due to the statistical uncertainty
in the mis-identification probability per track as a function of momentum.

For charged $K$'s  the data do not provide
a statistically rich and clean sample as for pions.  The
cleanest sample of charged kaons comes from reconstructed
$D^0 \rightarrow K^- \pi^+ (\pi^0)$ decays.  With $19742 \pm 221$
reconstructed $D^0$'s with a $K^-$ that passed the momentum cuts,
$4.5 \pm 5.5$ were consistent with the $K^-$ being identified as an electron.
This yields a central value of $(0.02 \pm 0.03) \%$  for the mis-identification
probability due to kaons.  As no momentum dependence measurement is possible
we use $(0.02 \pm 0.03) \%$ as the mis-identification probability for charged
kaons over the whole momentum range of interest.

Multiplying these mis-identification probabilities by the $\pi~:~K$
fractions, we find the total mis-identification probability of
$f_{\pi^+e^+}(X h^+)= (0.099 \pm 0.016) \%$ for the
right sign hadronic tracks and
$f_{\pi^+e^-}(X h^-) = (0.051 \pm 0.016) \%$
for the wrong sign hadronic tracks.  This is almost a factor of two
difference between the two mis-identification rates.  These
mis-identification probabilities represent the rate per
hadronic track from
$D^{*+} \rightarrow D^0 \pi^+$ decays where the $\pi^+$ had momentum
between 225 and 425 MeV/c.  Since the extraction of yields is done
in eight 25 MeV/c momentum bins, these mis-identification probabilities
are determined for each of the eight bins individually.  Small
variations arise due to different $D^0$ momentum
spectra and small changes in the $~\pi:~K$ ratio.

To turn these mis-identification probabilities into the actual yield of
mis-identified tracks, the inclusive right sign and
wrong sign rate
[$N(X h^+)$ and
$N(X h^-)$] is determined from the data.
The number of wrong sign and
right sign
hadronic
tracks associated with $D^{*+} \rightarrow D^0 \pi^+$ decays is determined
by using the same code and technique as for identified electrons, with the
requirement that the hadronic track not be identified as an electron.
The resulting estimated
mis-identified charged track contribution to the right and wrong sign
yields is given in Table~\ref{table:back_sum} as well as the final estimated
total background to the right sign yield.

\begin{table}
\begin{center}
\begin{tabular}{lllll}
$p(\pi)$    & $N_{ws}$
            & $F_{h^-}$
            & $F_{h^+}$
            & $N_{rs}$ \\
(MeV/c)     &
            &
            &
            & \\ \hline
225 - 250 & $32\pm 31$ & $13\pm3$ & $24\pm3$ & $43\pm31$\\
250 - 275 & $74\pm 29$ & $13\pm3$ & $22\pm3$ & $83\pm29$\\
275 - 300 & $45\pm 25$ & $9\pm2$ & $17\pm2$ & $53\pm25$\\
300 - 325 & $39\pm 22$ & $7\pm2$ & $13\pm2$ & $45\pm22$\\
325 - 350 & $-29\pm 18$& $5\pm1$ & $9\pm1$ & $-25\pm18$\\
350 - 375 & $36\pm 17$ & $4\pm1$ & $6\pm1$ & $38\pm17$\\
375 - 400 & $-4\pm 12$ & $2\pm1$ & $4\pm1$ & $-2\pm12$\\
400 - 425 & $0 \pm 11$ & $1\pm1$ & $2\pm1$ & $1\pm11$\\ \hline
Total         & $193\pm 62$& $54\pm5$ & $97\pm6$ & $236\pm64$\\ \hline
\end{tabular}
\caption{Summary of the expected background contribution as a function
of pion momentum to
the right sign yield, where $N_{rs} = N_{ws} - F_{h^-} + F_{h^+}$,
$F_{h^-} = N(Xh^-)f_{\pi^+e^-}(Xh^-)$ and
$F_{h^+} = N(Xh^+)f_{\pi^+e^+}(Xh^+)$.
\label{table:back_sum}
}
\end{center}
\end{table}

\section{Efficiency}

The
efficiency for detecting the $e^+$ determined by the
Monte Carlo simulation depends on the cocktail of exclusive modes
used to generate the inclusive semi-electronic decays.
The ratios of exclusive
rates presented in Appendix A are used to calculate the
ratios
$X_m = {\cal B}(D^0 \rightarrow m e^+ \nu)/
\sum_n {\cal B}(D^0 \rightarrow n e^+\nu)$,
where $m,n = K^-,  K^{*-}, K_1^-(1270),
K^{*-}(1430), \pi^-$, and $\rho^-$ mesons.
The efficiency for each of these modes is obtained from a Monte
Carlo simulation of each individual mode.  The inclusive efficiency
is obtained from
\begin{eqnarray}
\epsilon(Xe^+\nu) & = & \sum_m X_m \times \epsilon(D^0 \rightarrow m e^+ \nu).
\end{eqnarray}

The extraction of yields is done in eight pion  momentum bins from
225 to 425 MeV/c, as in the $D^0 \rightarrow K^- \pi^+$ analysis.
Table~\ref{table:final_yields} contains the efficiency in each
of the eight pion momentum bins.  The efficiencies for the
individual exclusive channels are in Table~\ref{table:eff_excl_all}
(Appendix A).
The total systematic error
due to uncertainties in the cocktail is determined by varying the ratios
in Table~\ref{table:excl} by one standard deviation, individually and
collectively.
The largest variation in the overall efficiency is seen when
$X_K$ and $X_\pi$ are both raised or both lowered and the other modes are
changed in the opposite direction.
This causes a $\pm 2\%$ change in the efficiency and is the estimated
systematic error due to the uncertainties in the cocktail of exclusive modes.

In addition to changing the cocktail the effect of the assumed
$q^2$ dependence
of the form factors is studied by changing the ISGW slope ($\kappa$)
{}~\cite{isgw2}.
The value used to generate the decays is $\kappa = 0.57 \pm 0.07$,
measured in a large sample of $D^0 \rightarrow K^- e^+\nu$ decays
by CLEO
{}~\cite{cleo_93}. Variations of one
sigma on $\kappa$ resulted in a $\pm0.6\%$ variation in efficiency.
The longitudinal and transverse contributions from
$D^0 \rightarrow K^{*-}e^+\nu$ decays were varied by
one sigma of their measured value and the total efficiency
changed by less than $\pm0.08\%$ \cite{kstarff}.

\section{Results}
\subsection{${\cal B} (D^0 \rightarrow X e^+ \nu)$}
The relevant measurements for determining ${\cal B}(D^0 \rightarrow Xe^+\nu)$
are given in Table~\ref{table:final_yields}.
The first column gives
the inclusive
$D^{*+} \rightarrow D^0 \pi^+$ yields from Ref.~\cite{cleo_kpi_93},
the second gives the background subtracted yield of $D^0 \rightarrow Xe^+\nu$
decays, followed by a column of efficiencies.  The last column is the
branching fraction for $D^0 \rightarrow Xe^+\nu$ for the eight momentum
bins.
As a check that the eight measurements are self consistent, the $\chi^2$
was calculated under the assumption that all
eight branching fraction measurements come from the weighted average.
The result is a $\chi^2$ of 9.4 for 7 degrees of
freedom.

Sources of
systematic effects and their estimated magnitude
are listed in Table~\ref{table:sys}.
The dominant systematic uncertainty is the evaluation of the electron
identification efficiency.  The electron identification algorithm was
developed using clean radiative Bhabha events in the data sample.  Its
performance on continuum events is studied using
$\pi^0\rightarrow \gamma e^+e^-$ where the
$e^+e^-$ pair could originate from either a Dalitz decay of the $\pi^0$
or a $\gamma$ conversion in material.  This study resulted in a conservative
estimate of the electron identification systematic uncertainty of $\pm3\%$.

The inclusive semi-electronic branching fraction is measured
to be
\begin{eqnarray}
{\cal B} (D^0 \rightarrow X e^+ \nu) = [6.64 \pm 0.18 \pm 0.29] \%
\end{eqnarray}
where the first error is statistical and the second error is the
estimated systematic effect.   Sources of model dependence have been
minimized by relying on the experimental measurements of the
exclusive rates of the observed modes and
experimental measurements of the  $d\Gamma/dq^2$ spectrum
in $D^0 \rightarrow K^-e^+\nu$ decays.  Models have been used only
for the $d\Gamma/dq^2$ spectrum of the other exclusive modes.
The previous value of $[7.01 \pm 0.62] \%$ agrees with this result~\cite{pdg}.

\begin{table}
\begin{center}
\begin{tabular}{lrrrr}
$p(\pi)$    & $N(D^{*+}\rightarrow D^0\pi^+$)
            & $N(D^{*+}\rightarrow D^0\pi^+$,
            & $\epsilon(Xe^+\nu)$
            & ${\cal B} (D^0 \rightarrow X e^+ \nu)$\\
(MeV/c)     &
            &  $D^0\rightarrow Xe^+\nu$)
            &  (\%)
            &  (\%)\\ \hline
225 - 250 & $44161\pm611$ &$1189\pm61$ & $37.9$ & $7.10\pm0.38$\\
250 - 275 & $39114\pm562$ & $988\pm57$ & $40.1$ & $6.30\pm0.38$\\
275 - 300 & $29482\pm475$ & $882\pm51$ & $42.7$ & $7.01\pm0.42$\\
300 - 325 & $21120\pm396$ & $644\pm44$ & $43.7$ & $6.97\pm0.49$\\
325 - 350 & $14973\pm334$ & $439\pm37$ & $45.5$ & $6.42\pm0.56$\\
350 - 375 & $ 9165\pm267$ & $221\pm30$ & $48.0$ & $5.02\pm0.70$\\
375 - 400 & $ 5492\pm208$ & $168\pm23$ & $49.5$ & $6.18\pm0.88$\\
400 - 425 & $ 2151\pm147$ & $ 78\pm19$ & $50.7$ & $7.15\pm 1.8$\\ \hline
Total         &$165658\pm1149$&$4609\pm121$&       & $6.64\pm0.18$\\ \hline
\end{tabular}
\caption{The yields of inclusive $D^{*+}\rightarrow D^0\pi^+$ decays,
$D^0\rightarrow Xe^+\nu$ decays, the efficiency for detecting the $Xe^+\nu$
final state and the calculated branching fraction, as a function of
the initial $D^{*+}$ pion momentum.  The errors
are statistical only and include the statistical error on the background
subtraction.
\label{table:final_yields}
}

\end{center}
\end{table}
\begin{table}
\begin{center}
\begin{tabular}{lc}
Source & Estimated Systematic Error \\
       &    ($\%$)              \\ \hline
Electron Identification Efficiency    & $\pm3.0$   \\
$Xe^+\nu$ Cocktail     & $\pm2.0$    \\
$N(D^{*+})$            & $\pm1.5$    \\
Track Reconstruction   & $\pm1.0$    \\
Monte Carlo Statistics & $\pm1.0$    \\
Electron fake rate     & $\pm1.0$    \\
Form Factor slope $\kappa$              & $\pm0.6$    \\ \hline
Total                  & $\pm4.3$  \\ \hline
\end{tabular}
\caption{Estimate of the systematic uncertainty in the
measurement of ${\cal B}(D^0 \rightarrow X e^+ \nu)$
\label{table:sys}
}
\end{center}
\end{table}

\subsection{ ${\cal B} (D^0 \rightarrow X e^+ \nu)/
               {\cal B} (D^0\rightarrow K^-\pi^+)$}
In addition to measuring the absolute $D^0 \rightarrow Xe^+\nu$ branching
fraction, it is straightforward to combine the yields presented here
with those in Ref.~\cite{cleo_kpi_93} to obtain a measurement of
the ratio
${\cal B}(D^0\rightarrow Xe^+\nu)/{\cal B}(D^0\rightarrow K^-\pi^+)$.
This tabulation
is done in Table~\ref{table:xenu_kpi}.  This ratio is independent of
systematics associated with the inclusive $D^{*+} \rightarrow D^0\pi^+$
yields.
The contributions to the systematic
error are given in Table~\ref{table:sys_ratio}.
The result is
\begin{eqnarray}
{\cal B} (D^0 \rightarrow X e^+ \nu)/
{\cal B} (D^0 \rightarrow K^- \pi^+) = 1.684 \pm 0.056 \pm 0.093.
\label{eq:xenukpi}
\end{eqnarray}
Again the first error is statistical and the second error is the
estimated systematic effect, where the use of a common dataset
allowed cancelation of some systematic effects present in the
individual results.

\begin{table}
\begin{center}
\begin{tabular}{lrrrrr}
$p(\pi)$    & $N(D^{*+}\rightarrow D^0\pi^+,$
            & $\epsilon(K\pi)$
            & $N(D^{*+}\rightarrow D^0\pi^+,$
            & $\epsilon(Xe^+\nu)$
            & $\frac{{\cal B} (D^0 \rightarrow X e^+ \nu)}
                    {{\cal B} (D^0\rightarrow K^-\pi^+)}$\\
(MeV/c)     & $D^0\rightarrow K^-\pi^+)$
            &  (\%)
            &  $D^0\rightarrow X e^+ \nu)$
            &  (\%)
            &  \\ \hline
225 - 250 & $1129\pm44$ &64.6 &$1189\pm61$ & $37.9$ & $1.80\pm0.12$\\
250 - 275 & $ 945\pm40$ &64.3 & $988\pm57$ & $40.1$ & $1.68\pm0.12$\\
275 - 300 & $ 741\pm34$ &64.4 & $882\pm51$ & $42.7$ & $1.80\pm0.13$\\
300 - 325 & $ 528\pm30$ &65.1 & $644\pm44$ & $43.7$ & $1.82\pm0.16$\\
325 - 350 & $ 393\pm25$ &66.0 & $439\pm37$ & $45.5$ & $1.62\pm0.18$\\
350 - 375 & $ 262\pm19$ &66.4 & $221\pm30$ & $48.0$ & $1.17\pm0.18$\\
375 - 400 & $ 153\pm15$ &68.8 & $168\pm23$ & $49.5$ & $1.53\pm0.26$\\
400 - 425 & $  57\pm9 $ &63.1 & $ 78\pm19$ & $50.7$ & $1.70\pm0.50$\\ \hline
Total         &$ 4208\pm83$ &     & $4609 \pm121$&       & $1.684\pm0.056$\\
\hline
\end{tabular}
\caption{The yields of $D\rightarrow K^-\pi^+$ decays,
the efficiency for detecting the $K^-\pi^+$ final state,
the yields of $D^0\rightarrow Xe^+\nu$ decays,
the efficiency for detecting the $Xe^+\nu$
final state and the calculated ratio of branching fractions, as a function of
the initial $D^{*+}$ pion momentum.  The errors on the data yields
are statistical only.
The ratio of branching fractions error is statistical only.
\label{table:xenu_kpi}
}
\end{center}
\end{table}
\begin{table}
\begin{center}
\begin{tabular}{lc}
Source & Estimated Systematic Error \\
       &    ($\%$)              \\ \hline
Electron Efficiency        &  $\pm3.0$   \\
$Xe^+\nu$ Cocktail         &  $\pm2.0$    \\
Track Reconstruction       &  $\pm3.8$    \\
Monte Carlo Statistics     &  $\pm1.2$    \\
Electron fake rate         &  $\pm1.0$    \\
Form Factor slope $\kappa$                  &  $\pm0.6$    \\
$K^-\pi^+$ (mass fit and  momentum cut)& $\pm0.7$ \\ \hline
Total                  &  $\pm5.5$  \\ \hline
\end{tabular}
\caption{Estimate of the systematic uncertainty in the measurement
of ${\cal B}(D^0 \rightarrow Xe^+\nu)/{\cal B}(D^0 \rightarrow K^- \pi^+)$.
\label{table:sys_ratio}
}
\end{center}
\end{table}

This ratio
allows for a check
of the ratio
\begin{eqnarray}
X_K& = &{\cal B}(D\rightarrow K^-e^+\nu)/{\cal B}(D\rightarrow Xe^+\nu) \\
   & = &\frac{{\cal B}(D\rightarrow K^-e^+\nu)}
             {{\cal B}(D\rightarrow K^-\pi^+)} \times
        \frac{{\cal B}(D\rightarrow K^-\pi^+)}
             {{\cal B}(D\rightarrow Xe^+\nu)}
\label{eq:e11}
\end{eqnarray}
which is used  in the $D^0 \rightarrow Xe^+\nu$ cocktail.
To obtain the most precise value possible, we take advantage of the fact that
the CLEO results for
${\cal B}(D^0 \rightarrow K^-e^+\nu)/{\cal B}(D^0\rightarrow K^-\pi^+)$
were obtained with the same detector, allowing reduction in
the systematic bias due to
lepton identification (reduced to $\pm1.7\%$) and the systematic bias
due to tracking reconstruction (reduced to $\pm2\%$).
There is also a large overlap of $D^0 \rightarrow K^- \pi^+$
events which were used to calculate the two
ratios which appear in Eq. ~\ref{eq:e11} ~\cite{cleo_reduced_errors}.
Using only CLEO results and taking these
common systematic effects into account
we obtain
$X_{K} = 0.581 \pm 0.023 \pm 0.028$.
Using all measurements of ${\cal B}(D^0 \rightarrow K^-e^+\nu)/
{\cal B}(D^0\rightarrow K^-\pi^+)$ and taking advantage of the
common CLEO systematic errors
results in a value of $X_{K} = 0.552 \pm 0.035$~\cite{correlation}.
These results agree well with
the input value of $X_{K}$ listed in Table~\ref{table:excl}.

\subsection{Comparison of inclusive measurement to the
sum of the exclusive rates}

The measurement of the inclusive semi-electronic branching fraction
is often compared to the sum of the measured
exclusive channels~\cite{bandwagon}.
This comparison provides a measure of the consistency of
the experimental measurements.

In terms of the branching fraction ratios,
$R_m = {\cal B}(D^0 \rightarrow m e^+ \nu)/
       {\cal B}(D^0\rightarrow K^- e^+\nu)$,
which are used in Appendix A for tabulating the $D^0 \rightarrow Xe^+\nu$
cocktail listed in Table~\ref{table:excl}, the ratio between the difference
of the inclusive and the sum of the exclusive rates can be written as:
\begin{eqnarray}
\frac{{\cal B}(D^0 \rightarrow Xe^+\nu) -
       \sum_m {\cal B}(D^0 \rightarrow m e^+ \nu)}
     {{\cal B}(D^0 \rightarrow Xe^+\nu)}
& = & 1-
X_K (1+R_{K^*}+R_{\pi}+R_\rho) \label{eq:sum_width2}.  \nonumber \\
 & &
\end{eqnarray}

Performing the  comparison  using only CLEO data
($X_K=0.581 \pm 0.036$,
and $1+R_{K^*}+R_\pi=1.724 \pm 0.078$) results in a value of:
\begin{eqnarray}
\frac{{\cal B}(D^0\rightarrow Xe^+\nu) -
\sum_m {\cal B}(D^0\rightarrow m e^+ \nu)}
     {{\cal B}(D^0\rightarrow Xe^+\nu)}
 & =& (-0.2 \pm 7.7) \%
{}.
\end{eqnarray}
This CLEO result does not include a contribution from $R_\rho$ as
CLEO has not reported a value for this ratio.  Inclusion of the small
contribution for $R_\rho$ will result in a central value further from zero,
while still entirely consistent with zero given the experimental errors.
Using the value of $X_K=0.552\pm 0.035$
obtained in the previous section and
$1+R_{K^*} +R_\pi+R_\rho = 1.751 \pm 0.067$
(see Table ~\ref{table:excl}) we find,
\begin{eqnarray}
\frac{{\cal B}(D^0\rightarrow Xe^+\nu) -
\sum_m {\cal B}(D^0\rightarrow m e^+ \nu)}
     {{\cal B}(D^0\rightarrow Xe^+\nu)}
 & =& (3.3 \pm 7.2) \%
{}.
\end{eqnarray}
These results
are consistent with the upper limits obtained by direct
searches for the unobserved exclusive modes~\cite{upperlimits}.

\subsection{The inclusive electron momentum spectrum}
The lepton spectrum from semi-leptonic charm decays has not been
updated since the DELCO results~\cite{delco}.  Because the measurement
presented here is not made in the rest frame of the $D^0$ we compare
the observed lepton spectrum in the lab frame with that of the Monte Carlo
simulation.
To obtain the momentum spectrum for inclusive $D^0 \rightarrow Xe^+\nu$
decays, events were selected if they pass all the selection criteria
previously described.  An additional cut
of  $\sin^2\alpha < 0.12$ is applied.  This cut
retains  $90\%$ of the signal and is large enough
that systematics associated with modeling the thrust axis
are minimized.  There is still background in this sample whose
shape is provided by the wrong sign $\sin^2\alpha$
distribution.  The normalization of this background is obtained
by normalizing the wrong sign $\sin^2\alpha$ distribution
to the right sign $\sin^2\alpha$ distribution for values of
$\sin^2\alpha >0.2$.
As this result is focused on the distribution of the electron momentum
not the normalization, a  $\pm1\%$ uncertainty
in the level of background
spread over the momentum range is negligible.
In Figure~\ref{figure:e_mom_spec}
the background subtracted momentum spectrum for
the electrons is shown along with the momentum spectrum obtained
from the Monte Carlo simulation.  The two distributions are normalized
to the same number of events, resulting in a 75\% confidence level.
The comparison
shows that the simulation is correctly producing $D^{*+}$, $D^0$
mesons and the inclusive $D^0 \rightarrow Xe^+\nu$ decays.
Any deviations would
indicate a problem in the simulation, either in the production or
decay dynamics.  We conclude that the
Monte Carlo provides a good simulation of  the data.

\begin{figure}
\vspace{6.0in}
 \caption{The lab momentum spectrum of electrons from semi-electronic
$D^0$ decays.  The solid squares represent the background subtracted
data, and the histogram is the result of a Monte Carlo simulation.
  \label{figure:e_mom_spec}}
\end{figure}

\section{Conclusions}
We have presented a new measurement of the inclusive branching
fraction for $D^0 \rightarrow Xe^+\nu$ decays.   The final result is,
\begin{eqnarray}
{\cal B} (D^0 \rightarrow X e^+ \nu)& = &
[6.64 \pm 0.18(stat.) \pm 0.29(syst.)] \%.
\end{eqnarray}

We find that the difference between this inclusive rate and the sum of the
observed exclusive channels is
$(3.3\pm7.2)\%$ of the inclusive rate.
This corresponds to an  upper limit
on the unobserved modes
of $14\%$ of the inclusive rate (at the $90\%$ C.L.).
The experimental upper limits obtained
using direct searches for specific
unobserved exclusive
semi-electronic modes are lower than the limit
quoted here.
However, the upper limit obtain in this paper is less
sensitive to the assumption of what exclusive channels are
unobserved.
The two methods, direct searches and inclusive-exclusive
rate comparison, both suggest that the remaining unobserved exclusive
semi-leptonic modes occur at small rates.
In addition the observed electron momentum spectrum
from inclusive $D^0 \rightarrow Xe^+\nu$ decays
is seen to be well described by the exclusive semi-electronic
cocktail.

\section{Acknowledgments}
We gratefully acknowledge the effort of the CESR staff in providing us with
excellent luminosity and running conditions.
J.P.A., J.R.P., and I.P.J.S. thank
the NYI program of the NSF,
G.E. thanks the Heisenberg Foundation,
%
%
K.K.G., M.S., H.N.N., T.S., and H.Y. thank the
OJI program of DOE,
J.R.P, K.H., and M.S. thank the A.P. Sloan Foundation,
and A.W., and R.W. thank the
Alexander von Humboldt Stiftung
for support.
This work was supported by the National Science Foundation, the
U.S. Department of Energy, and the Natural Sciences and Engineering Research
Council of Canada.

\appendix
\section{Determination of the $D^0 \rightarrow Xe^+\nu$ cocktail}
\label{appendix:excl}
In this appendix, the exclusive semi-leptonic branching fractions, and a list
of their averages are presented.
This list, which is referred to as the
$D^0 \rightarrow
Xe^+\nu$ cocktail, is used to calculate the electron detection efficiency.
The
$D^0 \rightarrow Xe^+\nu$
cocktail is determined using  world averages  to
obtain the following ratios:
\begin{eqnarray}
R_{K^*}&=&
{\cal B}(D^0 \rightarrow K^{*-} e^+ \nu)/
{\cal B}(D^0 \rightarrow K^- e^+ \nu) \\
R_{\pi}&=&
{\cal B}(D^0 \rightarrow \pi^- e^+ \nu)/
{\cal B}(D^0 \rightarrow K^- e^+ \nu) \\
R_{\rho}&=&
{\cal B}(D^0 \rightarrow \rho^- e^+ \nu)/
{\cal B}(D^0 \rightarrow K^- e^+ \nu).
\end{eqnarray}

Experimental upper limits are used to obtain estimates
for the unobserved modes:
\begin{eqnarray}
R_{K(1270)}& = &
{\cal B}(D^0 \rightarrow K_1^-(1270) e^+ \nu)/
{\cal B}(D^0 \rightarrow K^- e^+ \nu) \\
R_{K^*(1430)}& = &
{\cal B}(D^0 \rightarrow K^{*-}(1430) e^+ \nu)/
{\cal B}(D^0 \rightarrow K^- e^+ \nu).
\end{eqnarray}
The central value used for these unobserved modes is set to half the
$90\%$ confidence level upper limit and with an error equal to $\pm100\%$
of the central value.

The ratio of an exclusive channel to the inclusive rate is then obtained
from the following formulas:
\begin{eqnarray}
S & = & 1 + R_{K^*} + R_\pi + R_\rho +R_{K(1270)} + R_{K^*(1430)}
\label{eq:s}\\
X_K  & = &1/S \\
X_{K^*} & = & R_{K^*}/ S  \\
X_{\pi} & = & R_{\pi}/ S  \\
X_{\rho} & = & R_{\rho}/ S  \\
X_{K(1270)} & = & R_{K(1270)}/ S  \\
X_{K^*(1430)} & = & R_{K^*(1430)}/ S. \label{eq:x1430}
\end{eqnarray}

Throughout this appendix the
results are written in terms of the $D^0$ branching fractions.
Results from the $D^+$ sector are converted into $D^0$ equivalent branching
fractions using isospin and the measured $D^0$ and $D^+$ lifetimes.
Also semi-muonic measurements are converted into semi-electronic results
by correcting for the phase space difference between the muonic and electronic
modes~\cite{pdg}.
In several of the tables, two averages are presented,
one which includes all the data presented in the table, and
another with CLEO results excluded.  This is done to avoid
double weighting in the CLEO data when performing calculations.

\subsection{
$R_{K^*} = {\cal B}(D^0 \rightarrow K^{*-} e^+ \nu)$/
${\cal B}(D^0 \rightarrow K^- e^+ \nu)$}
There are two methods to measure this ratio: direct and indirect.
The direct measurements,  given in Table~\ref{table:kst_1},
can only be performed when both the $K$ and $K^*$ modes are
reconstructed through the same parent species within the same
experiment.  The indirect
measurement compares the $K^*e^+\nu$ width measured in $D^+$ decays
to the $K^-e^+\nu$ width measured in $D^0$ decays, via
\begin{eqnarray}
R^{indirect}_{K^*}  &  = &
    \frac{{\cal B}(D^+ \rightarrow {\bar K}^{*0} e^+ \nu)}
         {{\cal B}(D^+ \rightarrow K^- \pi^+ \pi^+)}\times  \nonumber \\
&&    \frac{{\cal B}(D^0 \rightarrow K^- \pi^+)}
         {{\cal B}(D^0 \rightarrow K^- e^+ \nu)}\times  \nonumber \\
&&    \frac{{\cal B}(D^+ \rightarrow K^- \pi^+ \pi^+)}
         {{\cal B}(D^0 \rightarrow K^- \pi^+)}  \times
                              \frac{\tau_{D^0}}{\tau_{D^+}}.
\end{eqnarray}
Table~\ref{table:kenu} contains the world average for
${\cal B}(D^0 \rightarrow K^- e^+ \nu)/
 {\cal B}(D^0 \rightarrow K^- \pi^+)$
and
Table~\ref{table:kst_2} contains the world average for
${\cal B}(D^+ \rightarrow {\bar K}^{*0} e^+ \nu)/
 {\cal B}(D^+ \rightarrow K^- \pi^+ \pi^+)$
where
the CLEO measurements have been specifically excluded as
these measurements are used in the direct determination of $R_{K^*}$.
To determine $R_{K^*}^{indirect}$,
the ratio of normalizing modes $K\pi\pi/K\pi$
presented in Table~\ref{table:kpi} is used.
Using the world average for
this ratio of branching fractions and the $D^+/D^0$
lifetime ratio \cite{pdg} the value for $R^{indirect}_{K^*}$ is
measured to be
$0.559 \pm 0.068$.  Averaging $R^{direct}_{K^*}$ and $R^{indirect}_{K^*}$
yields
\begin{eqnarray}
R_{K^*} = 0.579 \pm 0.049.
\end{eqnarray}

\begin{table}
\begin{center}
\begin{tabular}{lcr}
Experiment    & Reference        & ${\bar K}^*e^+\nu/{\bar K}e^+\nu$  \\ \hline
CLEO93        & \cite{cleo_93}   & $0.62 \pm 0.08$  \\
CLEO91        & \cite{cleo_91}   & $0.51 \pm 0.19$  \\ \hline
Average       &                  & $0.60 \pm 0.07$   \\  \hline
\end{tabular}
\caption{Direct measurements of the
${\cal B}(D \rightarrow K^* e^+ \nu)/{\cal B}(D \rightarrow K e^+ \nu)$ ratio
and their weighted average.\label{table:kst_1}
}
\end{center}
\end{table}
\begin{table}
\begin{center}
\begin{tabular}{lcr}
Experiment &  Reference          &  $K^-e^+\nu/K^-\pi^+$  \\ \hline
CLEO93       & \cite{cleo_93}      &  $0.978 \pm 0.052$  \\
E687 (94)  & \cite{e687_kenu_94} &  $0.865 \pm 0.051$ \\
CLEO91     & \cite{cleo_91}      &  $0.86 \pm 0.07$ \\
E691       & \cite{e691_kenu}    &  $0.91 \pm 0.13$ \\
E687 (90)  & \cite{e687_kenu_90} &  $0.84 \pm 0.19$ \\ \hline
Average without CLEO   &         &  $0.869 \pm 0.046$ \\
Average    &                     &  $0.906 \pm 0.031$ \\ \hline
\end{tabular}
\caption{Measurements of the
${\cal B}(D^0\rightarrow K^- e^+ \nu)/{\cal B}(D^0 \rightarrow K^- \pi^+)$
ratio and
their weighted average.  The average without  CLEO measurements
is also calculated separately to avoid multiple use of the CLEO results
in determining $R_{K^*}$.
\label{table:kenu}}
\end{center}
\end{table}
\begin{table}
\begin{center}
\begin{tabular}{lcr}
Experiment    & Reference        & ${\bar K}^{*0}e^+\nu/K^-\pi^+\pi^+$  \\
\hline
E691          & \cite{e691_kst}  & $0.49 \pm 0.06$    \\
E687          & \cite{e687_kst}  & $0.59 \pm 0.07$   \\
CLEO          & \cite{cleo_93}   & $0.67 \pm 0.11$  \\
E653          & \cite{e653_kst}  & $0.48 \pm 0.11$   \\
Argus         & \cite{argus_kst} & $0.55 \pm 0.13$   \\
WA82          & \cite{wa82_kst}  & $0.62 \pm 0.17$   \\ \hline
average       & without CLEO     & $0.527 \pm 0.041$ \\
average       &                  & $0.547 \pm 0.038$  \\  \hline
\end{tabular}
\caption{Measurements of the
${\cal B}(D^+ \rightarrow {\bar K}^{*0} e^+ \nu)/
 {\cal B}(D^+ \rightarrow K^- \pi^+ \pi^+)$ ratio
and their weighted average.\label{table:kst_2}
}
\end{center}
\end{table}
\begin{table}
\begin{center}
\begin{tabular}{lcccc}
Experiment  & Reference
& ${\cal B}(D^0 \rightarrow K^- \pi^+)$
& ${\cal B}(D^+ \rightarrow K^- \pi^+ \pi^+)$
& $\frac{{\cal B}(D^+ \rightarrow K^- \pi^+ \pi^+)}
        {{\cal B}(D^0 \rightarrow K^-\pi^+)}$ \\
  &
& (\%)
& (\%)
&  \\ \hline
CLEO        & \cite{cleo_kpi_93}   & $3.91 \pm 0.19$ &$9.3 \pm 1.0$
& $2.35 \pm 0.23$\\
ARGUS       & \cite{argus_kpi_94}  & $3.41 \pm 0.30$ &&\\
ALEPH       & \cite{aleph_kpi}     & $3.89 \pm 0.33$ && \\
Mark III    & \cite{mark3_kpi}     & $4.2 \pm 0.6$   &$9.1 \pm 1.4$  & \\
Mark II     & \cite{mark2_kpi}     & $4.1 \pm 0.6$   &$9.1 \pm 1.9$  & \\
ARGUS       & \cite{argus_kpi_93}  & $4.5 \pm 0.7$   &&\\
HRS         & \cite{hrs_kpi}       & $4.50 \pm 0.94$ &&\\
Mark I      & \cite{mark1_kpi}     & $4.3 \pm 1.0$   &$8.6 \pm 2.0$  & \\
\hline
average     & without CLEO         & $3.84 \pm 0.18$ &$8.98 \pm 0.98$
& $2.34 \pm 0.28$ \\ \hline
average     &                      & $3.87 \pm 0.13$ &$9.1 \pm 0.7$
& $2.35 \pm 0.18$\\  \hline
\end{tabular}
\caption{Measurements of the hadronic normalizing
modes, $D^0 \rightarrow K^-\pi^+$, $D^+ \rightarrow K^- \pi^+ \pi^+$ and
their ratio.  The CLEO result on
${\cal B}(D^+ \rightarrow K^- \pi^+ \pi^+)/
{\cal B}(D^0 \rightarrow K^- \pi^+)$ is a direct measurement of this
ratio, and is not obtained by dividing the individual CLEO results.
\label{table:kpi}
}
\end{center}
\end{table}

\subsection{$R_\pi = {\cal B}(D^0 \rightarrow \pi^- e^+ \nu)/
                     (D^0 \rightarrow K^- e^+ \nu)$}
The Cabibbo suppressed decay $D^0 \rightarrow \pi^- e^+ \nu$ has been observed
at Mark III.
CLEO  has made measurements of both the $D^0 \rightarrow \pi^- e^+ \nu$
and the  $D^+ \rightarrow \pi^0 e^+ \nu$ decay chains.
There is factor of two due to
isospin that is needed to convert the $D^+ \rightarrow \pi^0 e^+ \nu$
measurement
to a $D^0 \rightarrow \pi^- e^+ \nu$ branching fraction.
The results are presented in Table~\ref{table:pienu}.
\begin{table}
\begin{center}
\begin{tabular}{lccc}
Experiment    & Reference     &Mode    & $\pi^- e^+ \nu/K^-e^+\nu$  \\ \hline
CLEO        & \cite{cleo_94_pienu}&$\pi^- e^+\nu$  & $0.103 \pm 0.041$  \\
Mark III      & \cite{mark3_pienu}  &$\pi^-e^+\nu$ & $0.115 \pm 0.051$    \\
CLEO        & \cite{cleo_93_pienu}&$\pi^0e^+\nu$ & $0.17 \pm 0.06$  \\ \hline
average       &        &            & $0.121 \pm 0.028$   \\ \hline
\end{tabular}
\caption{Measurements of the
${\cal B}(D^0 \rightarrow \pi^- e^+ \nu)/
 {\cal B}(D^0 \rightarrow K^- e^+ \nu)$ ratio
and their weighted average.\label{table:pienu}
}
\end{center}
\end{table}

\subsection{$R_\rho = {\cal B}(D^0 \rightarrow \rho^- e^+ \nu)/
                     {\cal B}(D^0 \rightarrow K^- e^+ \nu)$}
Fermilab experiment
E653 has published an observation  of four
$D^+ \rightarrow \rho^0 \mu^+ \nu$
events based on a kinematic separation of the Cabibbo suppressed
$\rho^0 \mu^+ \nu$ signal from the more copious
${\bar K}^{*0} \mu^+ \nu$ mode~\cite{e653_rho}.
They measure
${\cal B}(D^+ \rightarrow \rho^0 \mu^+ \nu)/
 {\cal B}(D^+ \rightarrow {\bar K}^{*0} \mu^+ \nu)
=
0.044^{+0.031}_{-0.025} \pm 0.014$. To obtain $R_\rho$ this measurement
needs be corrected by the isospin factor and multiplied by
$R_{K^*}$ which gives;
$R_\rho= {\cal B}(D^+ \rightarrow \rho^0 \mu^+ \nu)/
         {\cal B}(D^+ \rightarrow {\bar K}^{*0} \mu^+ \nu)
\times R_{K^*} \times I_\rho =
(0.044^{+0.031}_{-0.025} \pm 0.014) \times (0.579\pm 0.049)\times 2
= 0.051 \pm 0.037$.
For Monte Carlo generation it is assumed that the form factor ratios for
$D^0 \rightarrow \rho^- e^+ \nu$ decay are
identical to that of the well measured
$D^0 \rightarrow K^{*-} e^+ \nu$ decay.

\subsection{${\cal B}(D^0 \rightarrow ({\bar K}^* \pi)^- e^+ \nu)$
upper limits}
Searches for higher $K^{(*)}$ resonances and possible
non-resonant contributions
to $D$ semi-leptonic decay have been performed by the fixed target experiments
{}~\cite{upperlimits}.
Although no evidence for these
decays has been demonstrated  we include
$D^0 \rightarrow K^{-}(1270) e^+ \nu$ and
$D^0 \rightarrow K^{*-}(1430) e^+ \nu$
in the Monte Carlo simulation.  The decays are generated unpolarized and
with the following strengths and  errors,
$R_{K(1270)} =
 {\cal B}(D^0 \rightarrow K_1^-(1270) e^+ \nu)/
 {\cal B}(D^0 \rightarrow K^- e^+ \nu) = 0.03 \pm 0.03$
and
$R_{K^*(1430)} =
 {\cal B}(D^0 \rightarrow K^{*-}(1430) e^+ \nu)/
 {\cal B}(D^0 \rightarrow K^- e^+ \nu) = 0.02 \pm 0.02$.
It is assumed that any non-resonant contribution to the inclusive rate
will have a similar electron momentum spectrum  distribution
as these higher order modes.

\subsection{Calculation of the $D^0 \rightarrow Xe^+\nu$ cocktail}
Table~\ref{table:excl} summarizes the
relative rates, $R_m$ (relative to
$D^0 \rightarrow K^-e^+\nu$) obtained in the previous
sections.  The sum of these rates is then used to determine
the ratio of each exclusive rate to the sum of all the exclusive rates as per
Eqs.~\ref{eq:s}-~\ref{eq:x1430}.  Table~\ref{table:eff_excl_all} contains
the efficiencies for these exclusive modes to pass the selection
criteria.

\begin{table}
\begin{center}
\begin{tabular}{lcc}
Mode    &  $R_m$ & $X_m$ \\ \hline
$D^0 \rightarrow K^- e^+ \nu$      & 1.0               & $ 0.555 \pm 0.024$ \\
$D^0 \rightarrow K^{*-}e^+\nu$      &  $0.579 \pm 0.049$& $ 0.321 \pm 0.021$ \\
$D^0 \rightarrow \pi^- e^+ \nu$    &  $0.121 \pm 0.028$& $ 0.067 \pm 0.015$ \\
$D^0 \rightarrow \rho^- e^+ \nu$   &  $0.051 \pm 0.037$& $ 0.028 \pm 0.020$ \\
$D^0 \rightarrow K_1^-(1270)e^+\nu$  & $0.03 \pm 0.03$   & $ 0.017 \pm 0.016$
\\
$D^0 \rightarrow K^{*-}(1430)e^+\nu$&$0.02 \pm 0.02$ & $ 0.011 \pm 0.011$ \\
\hline
Sum            &$1.801 \pm 0.077$&     \\ \hline
\end{tabular}
\caption{The world average or estimate of the ratio of exclusive
channels relative to the $D^0 \rightarrow K^-e^+\nu$ decay mode,
$R_m= {\cal B}(D^0 \rightarrow m e^+\nu)/{\cal B}(D^0 \rightarrow K^-e^+\nu)$.
The third column, is the ratio of the exclusive rate to the sum of the
exclusive rates,
$X_m= {\cal B}(D^0 \rightarrow m e^+\nu)/$Sum.
\label{table:excl}
}
\end{center}
\end{table}
\begin{table}
\begin{center}
\begin{tabular}{ccccccc}
$p(\pi)$ &
$\epsilon(K^-e^+\nu)$ &
$\epsilon(K^{*-}e^+\nu)$ &
$\epsilon(\pi^- e^+\nu)$ &
$\epsilon(\rho^- e^+ \nu)$ &
$\epsilon(K_1^-(1270) e^+ \nu)$ &
$\epsilon(K^{*-}(1430) e^+ \nu)$ \\
MeV/c  &$(\%)$&$(\%)$&$(\%)$&$(\%)$&$(\%)$&$(\%)$ \\ \hline
225-250& 40.4 & 34.4 & 42.4 & 38.1 & 20.4 & 10.9 \\
250-275& 42.8 & 36.3 & 45.4 & 39.5 & 22.9 & 11.7 \\
275-300& 45.6 & 38.6 & 47.6 & 42.6 & 20.6 & 12.0 \\
300-325& 46.2 & 40.4 & 49.2 & 43.8 & 23.4 & 13.0 \\
325-350& 48.6 & 41.0 & 51.0 & 46.6 & 27.2 & 12.0 \\
350-375& 50.8 & 44.1 & 54.6 & 43.7 & 30.7 & 14.6 \\
375-400& 51.9 & 46.1 & 56.6 & 48.2 & 29.4 & 19.5 \\
400-425& 53.9 & 45.2 & 57.7 & 57.4 & 21.4 & 34.3 \\ \hline
\end{tabular}
\caption{Efficiencies for the exclusive decay channels used in the
$Xe^+\nu$ cocktail.
\label{table:eff_excl_all}
}
\end{center}
\end{table}

\subsection{Comparison of the inclusive rate to the sum of the exclusive
measurements.}

One of the most frequent comparisons in the literature
{}~\cite{jeffandpat,pdg,bandwagon}
is the sum of the observed exclusive channels to the measured
inclusive rate.
The method of comparing the inclusive measurement to
the sum of the ratio of exclusive measurements is
presented here.

The following set of equations are used to calculate the
branching fraction for the observed exclusive decays:
\begin{eqnarray}
{\cal B}(D^0 \rightarrow K^- e^+ \nu)            & = &
r^{Ke^+\nu}_{K\pi}\times {\cal B}(D^0 \rightarrow K^-\pi^+)
\label{eq:k_width}\\
{\cal B}(D^0 \rightarrow K^{*-} e^+ \nu) & = &
r^{Ke^+\nu}_{K\pi} \times {\cal B}(D^0 \rightarrow K^-\pi^+) \times R_{K^*}
\label{eq:kstar_width}\\
{\cal B}(D^0 \rightarrow \pi^- e^+ \nu) & = &
r^{Ke^+\nu}_{K\pi} \times {\cal B}(D^0 \rightarrow K^-\pi^+) \times R_{\pi} \\
{\cal B}(D^0 \rightarrow \rho^- e^+ \nu) & =  &
r^{Ke^+\nu}_{K\pi} \times {\cal B}(D^0 \rightarrow K^-\pi^+)\times R_{\rho}
\label{eq:rho_width}
\end{eqnarray}
The sum of the observed exclusive rates is then:
\begin{eqnarray}
\sum_m{\cal B}(D^0 \rightarrow m e^+ \nu)            & = &
r^{Ke^+\nu}_{K\pi}\times
{\cal B}(D^0 \rightarrow K^-\pi^+) \times \nonumber \\
 & & (1+R_{K^*}
+R_{\pi}+R_{\rho}) \label{eq:sum_width}
\end{eqnarray}
The quantities
$r^{Ke^+\nu}_{K\pi} = {\cal B}(D^0 \rightarrow K^-e^+\nu)/
{\cal B}(D^0 \rightarrow K^-\pi^+)$ and
${\cal B}(D^0 \rightarrow K^-\pi^+)$  are common to
all derived exclusive branching fractions, and thereby effect the entire scale.


\begin{thebibliography}{99}
\bibitem{conjugate}
Charge conjugation is implied throughout this paper, the word electron
will be used to refer to both $e^-$'s and $e^+$'s.
\bibitem{rareD}
Other sources of right and wrong sign electrons are $D^0 - {\bar D^0}$
mixing and flavor changing neutral currents, $D^0 \rightarrow e^+e^-$.
Experimental limits and theoretical expectations
on these rare processes make their contribution to the right
and wrong sign yields negligible~\cite{pdg}.
\bibitem{kstarff}
The $D \rightarrow {\bar K}^* e^+ \nu$ decay has three helicity states in the
limit of negligible lepton mass.  In Monte Carlo generation
the world average~\cite{e691_kst,e687_kst,e653_kst} of the three
form factors involved in the decay is used to determined the
ratio of longitudinal to transverse alignments
($\Gamma_L/\Gamma_T = 1.23 \pm 0.13$).
\bibitem{cleo_reduced_errors}
When combining the CLEO result for
${\cal B}(D^0 \rightarrow K^-e^+\nu)/
{\cal B}(D^0\rightarrow K^-\pi^+)$ and the CLEO result for
${\cal B} (D^0 \rightarrow X e^+ \nu)/
{\cal B} (D^0 \rightarrow K^- \pi^+)$ we use the following values
which have reduced systematic errors:
\begin{eqnarray}
{\cal B}(D^0 \rightarrow K^-e^+\nu)/
{\cal B}(D^0\rightarrow K^-\pi^+)&=&0.978\pm 0.025 \pm 0.032 \\
{\cal B} (D^0 \rightarrow X e^+ \nu)/
{\cal B} (D^0 \rightarrow K^- \pi^+)& =&
1.684 \pm 0.049 \pm 0.061.
\end{eqnarray}
\bibitem{correlation}
When combining the CLEO result for
${\cal B}(D^0 \rightarrow K^-e^+\nu)/
{\cal B}(D^0\rightarrow K^-\pi^+)$ and the CLEO result for
${\cal B} (D^0 \rightarrow X e^+ \nu)/
{\cal B} (D^0 \rightarrow K^- \pi^+)$ we use the following values
with reduced systematic errors:
\begin{eqnarray}
{\cal B}(D^0 \rightarrow K^-e^+\nu)/
{\cal B}(D^0\rightarrow K^-\pi^+)&=&0.978\pm 0.025 \pm 0.032 \\
{\cal B} (D^0 \rightarrow X e^+ \nu)/
{\cal B} (D^0 \rightarrow K^- \pi^+)& =&
1.684 \pm 0.049 \pm 0.061.
\end{eqnarray}
\bibitem{correlation}
This value for $X_K$ includes the CLEO and non-CLEO results for
$R={\cal B}(D^0 \rightarrow K^-e^+\nu)/{\cal B}(D^0 \rightarrow K^-\pi^+)$.
The weighted average of
$R(CLEO) = 0.978 \pm 0.041$ (from ~\cite{cleo_reduced_errors}) and
$R(non-CLEO) = 0.869 \pm 0.046$
is  $R(wtd)=0.930 \pm 0.031$.
To calculate $X_K$ one divides by the branching fraction
ratio $R' = {\cal B}(D^0 \rightarrow Xe^+\nu)/(D^0 \rightarrow K^-\pi^+)$.
Using only CLEO data to calculate $X_K$, the statistical and systematic
errors of $R$ and $R'$ are reduced because of the cancelation of the
contributions from ${\cal B}(D^0 \rightarrow K^- \pi^+)$, and
one calculates
$X_K = R/R' = (0.978 \pm 0.041)/(1.684 \pm 0.078)=0.581 \pm 0.036$,
where $R$ and $R'$ are taken from ~\cite{cleo_reduced_errors}.
Using $R(non-CLEO)$
to
calculate $X_K$, one obtains
$X_K = R/R' = (0.869 \pm 0.046)/(1.684 \pm 0.109)=0.516 \pm 0.043$.
In this case the full errors for $R'$ are used (see equation 9),
because the non-CLEO measurement of $R$ has no experimental correlation
with the CLEO measurement of $R'$.  To combine the CLEO and non-CLEO
results for $R$ to calculate $X_K$, one needs a weighted value of $R'$
from the two values given above.
$R'(wtd) = 1.684 \pm [0.078f(correlated) + 0.109f(uncorrelated)]$,
where $f(uncorrelated) = 1 - f(correlated)$ =
$(1/0.046)^2 /[(1/0.046)^2+(1/0.041)^2] = 0.438$.    This results
in $R'(wtd) = 1.684 \pm 0.092$.   Taking the
ratio
$R(wtd)/R'(wtd) =
(0.930 \pm 0.030)/(1.684 \pm 0.092)$ one calculates $X_K=0.552 \pm 0.035$.
\end{thebibliography}
\end{document}